\documentclass{article}

\usepackage{arxiv}
\usepackage{amsmath} 
\usepackage{longtable}
\usepackage{graphicx}
\usepackage[super,sort,comma]{natbib}
\usepackage[section]{placeins}

\usepackage[utf8]{inputenc} % allow utf-8 input
\usepackage[T1]{fontenc}    % use 8-bit T1 fonts
\usepackage{hyperref}       % hyperlinks
\usepackage{url}            % simple URL typesetting
\usepackage{booktabs}       % professional-quality tables
\usepackage{amsfonts}       % blackboard math symbols
\usepackage{nicefrac}       % compact symbols for 1/2, etc.
\usepackage{microtype}      % microtypography
\usepackage{lipsum}
\usepackage{graphicx}
\graphicspath{ {./images/} }

\title{Technical Note: Vendor-Specific Approach for Standardized Uptake Value Calculation}

\author{
 Maksym Fritsak \\
  Department of Radiation Oncology\\
  University Hospital and University of Zurich\\
  Zurich, Switzerland\\
  Faculty of Medicine, University of Zurich\\ 
  Zurich, Switzerland\\
  \texttt{Maksym.Fritsak@usz.ch} \\
   \And
 Hubert S. Gabryś \\
  Department of Radiation Oncology\\
  University Hospital and University of Zurich\\
  Zurich, Switzerland\\
  \texttt{Hubert.Gabrys@usz.ch} \\
  \And
 Preethi Mohan \\
  Department of Radiation Oncology\\
  University Hospital and University of Zurich\\
  Zurich, Switzerland\\ 
  Institute of Computational Life Sciences\\
  ZHAW School of Life Sciences and Facility Management\\
  Zurich, Switzerland\\
  \texttt{pritilatha@gmail.com} \\
   \And
 Matthias Guckenberger\\
  Department of Radiation Oncology\\
  University Hospital and University of Zurich\\
  Zurich, Switzerland\\
  Faculty of Medicine, University of Zurich\\ 
  Zurich, Switzerland\\
  \texttt{Matthias.Guckenberger@usz.ch} \\
  \And
 Stephanie Tanadini-Lang \\
  Department of Radiation Oncology\\
  University Hospital and University of Zurich\\
  Zurich, Switzerland\\
  \texttt{Stephanie.Tanadini-Lang@usz.ch} \\
}

\begin{document}
\maketitle
\begin{abstract}
\noindent {\bf Background:} The Standardized Uptake Value (SUV) is a critical metric in positron emission tomography (PET) imaging, used to assess metabolic activity. However, calculating SUV from DICOM files presents challenges due to vendor-specific DICOM attributes and variations in the encoding of radiotracer accumulation times.\\ 
{\bf Purpose:} This technical note introduces a robust, vendor-specific SUV calculation strategy designed to addresses inconsistencies in current methodologies. We also integrate this strategy into an open-source software, Z-Rad, capable of converting raw PET DICOM data into body-weight normalized SUV NIfTI files.\\
{\bf Methods:} Our SUV calculation strategy was developed by reviewing DICOM conformance statements. Validation was conducted using real-world PET datasets, and the proposed strategy was compared to existing commercial and open-source software tools.\\
{\bf Results:} Our proposed strategy demonstrated improved accuracy by resolving time-related discrepancies in the studied data, which can lead to substantial errors in SUV calculation. In particular, we identified that popular software solutions often overlook vendor-specific DICOM attributes and incorrect PET image series times, resulting in SUV estimation errors of up to 32\% in our test data. \\
{\bf Conclusions:} The proposed vendor-specific SUV calculation strategy facilitates quantitative PET imaging analysis by addressing variations in DICOM attributes encoding across different vendors. This method allows to avoid common mistakes with adherence to the DICOM conformance statements and has been integrated into an open-source software, Z-Rad.

\end{abstract}

% keywords can be removed

The standardized uptake value (SUV) is a fundamental quantitative metric in positron emission tomography (PET) \cite{Thie1431} to measure the relative concenration of a radiotracer in a specific region of interest providing a standardized way to assess metabolic activity. It has been shown to be predictive for several tumor types \cite{Lai2024, Shalaby2022}, essential for quantifying treatment response\cite{Vanderhoek1188}, and is a precursor to more comprehensive metrics like total lesion glycolysis (TLG) \cite{Kitao2016}, metabolic tumor volume (MTV) \cite{Kitao2016}, and PET-based radiomic studies. However, the methods for calculating SUV from DICOM files vary across vendors \cite{ge_healthcare_pet_dicom, philips_dicom_nuclear_medicine, siemens_dicom_molecular_imaging}, software solution\cite{slicer_suv_calculator}, and online challenges \cite{autopet_dataset}.

A major step toward standardizing SUV calculations was provided by Quantitative Imaging Biomarkers Alliance (QIBA) \cite{qiba_suv}. Paul Kinahan et al. surveyed four different vendors to develop both vendor-specific \cite{qiba_suv_pseudocode} and vendor-neutral \cite{qiba_suv_happypath} QIBA SUV calculation strategies. Their vendor-specific approach \cite{qiba_suv_pseudocode} takes into account private DICOM attributes from different vendors but only addresses errors when the \texttt{Series Time (0008,0031)} is after the \texttt{Acquisition Time (0008,0022)}. In contrast, their vendor-neutral “happy path only” strategy~\cite{qiba_suv_happypath} forgoes all vendor-specific tags and relies solely on \texttt{Series Time (0008,0031)}.

In this technical note, we present a refined, comprehensive vendor-specific SUV calculation strategy that addresses various pitfalls and challenges associated with real-world PET data. We provide a thorough analysis of existing SUV calculation methodologies, highlight potential issues, and propose a software solution to facilitate consistent, inter-vendor comparisons in PET imaging. Our objective is to improve quantitative reliability across a diverse range of DICOM datasets and scanning protocols.

\section{Methods}

\subsection{SUV Definition}

The standardized uptake value (SUV) normalized by the body weight, \(\text{SUV}_{bw}\), is typically defined as:
\begin{equation}
\label{eq:suv_def}
\text{SUV}_{bw} = \frac{A \cdot w}{D_i} \cdot 2^\frac{t_s-t_i}{T_{1/2}} \quad \left [ \frac{g}{mL} \right ],
\end{equation}
where \(D_i\) is the injected dose of the radiotracer (in Bq), \(w\) - patient's weight (in grams), \(A\) - activity concentration (in Bq/mL), \(t_s-t_i\)  - nonnegative difference between measurement and radiotracer injection times respectively (in seconds), and \(T_{1/2}\) - radiotracer half-life time (in seconds).

\subsection{Calculation of SUV from DICOM}
\subsubsection{Conversion of Stored Pixel Values to Physical Units}
In PET DICOM files, the \texttt{Rescale Slope  (0028,1053)} and \texttt{Intercept (0028,1052)} are used to convert the pixel values stored in the image into physical units, as defined by \texttt{Units (0054,1001)}. These units can represent measurements such as activity concentration in Becquerels per milliliter (BQML), grams per milliliter (GML), or counts (CNTS). A useful parameter for further discussion, \( A' \), can be derived from the DICOM file using the following equation:
\begin{equation}
\label{eq:dcm_intensity}
A' = P \cdot S + I,
\end{equation}
where \( P \) represents the original stored pixel value, \( S \) is the \texttt{Rescale Slope (0028,1053)}, and \( I \) is the \texttt{Rescale Intercept (0028,1052)}. The parameter \( A' \) is measured in the units specified in \texttt{Units (0054,1001)}. If units are BQML, then \( A' \) is equal to \( A \) from Eq. \ref{eq:suv_def}.

Most discrepancies in SUV calculation from DICOM files arise from incorrect handling of PET DICOM units by third-party software and differences in decay correction strategies. To address these issues, we derived a vendor-specific SUV calculation workflow based on the DICOM conformance statements from three major PET machine manufacturers: Philips \cite{philips_dicom_nuclear_medicine}, Siemens Healthineers \cite{siemens_dicom_molecular_imaging}, and GE \cite{ge_healthcare_pet_dicom}. Leveraging this information, we developed a unified framework to ensure consistency and accuracy in SUV definition across different systems (Figure \ref{SUV}). A comprehensive list of all used DICOM attributes is provided in the Supplement.

\subsubsection{Scan Start Time Estimation}
In metabolic PET imaging, the most commonly used \texttt{Units (0054,1001)} value is BQML. For this unit, when the \texttt{Decay Correction (0054,1102)} is set to ADMIN, all images are decay-corrected to the radiopharmaceutical injection time \(t_i\). This implies that \(t_s\) is equal to \(t_i\) in Eq. \ref{eq:suv_def}.

If the \texttt{Decay Correction (0054,1102)} value is set to START, the image frames are decay-corrected to the scan start time. In principle, this should align with \texttt{Series Time (0008,0031)}. However, because \texttt{Series Time} can be altered by subsequent processing, Siemens Healthineers and GE provide private DICOM tags (\texttt{(0071,1022)} for Siemens Healthineers and \texttt{(0009,100d)} for GE) to store a more reliable \texttt{Decay Correction Date Time}. 

If these private tags are missing, according to the GE DICOM Conformance Statement, the scan start time \(t_s\) can be reconstructed as following\cite{ge_healthcare_pet_dicom}:
\begin{equation}
\label{eq:ge_decay_time}
    t_s = t_{a} - \Delta t_{FRT},
\end{equation}
where $t_a$ is the \texttt{Acquisition Time (0008,0022)} and $\Delta t_{FRT}$ is the \texttt{Frame Reference Time(0054,1300)} value in seconds. This indicates, that the images are decay corrected to the start of a frame. Similarly for Siemens Healthineers, the scan start time can be calculated based on the information from their DICOM Conformance Statement. Specifically, the provided definition of the Decay Factor implies that images are decay corrected to the time at which the average activity in a frame is corrected (\(\Delta t_{AVE}\))\cite{siemens_dicom_molecular_imaging}:
\begin{equation}
\label{eq:siemens_decay_time}
    t_s = \Delta t_{a} + \Delta t_{AVE} - \frac{\ln(\kappa)}{\lambda},
\end{equation}
\begin{equation}
\label{eq:tave}
    \Delta t_{AVE} = \frac{1}{\lambda} \ln{\frac{\lambda \Delta t_{AFD}}{1-e^{-\lambda \Delta t_{AFD}}}},
\end{equation}
where $\Delta t_{AFD}$ is the \texttt{Actual Frame Duration (0018,1242)} value (in seconds), $\kappa$ is the \texttt{Decay Factor (0054,1321)}, and $\lambda$ = $\ln(2)/T_{1/2}$ is the decay constant, with $T_{1/2}$ specified in the \texttt{Radiopharmaceutical Information Sequence (0054,0016)}. Since Siemens Healthineers implementation also sets \texttt{Frame Reference Time(0054,1300)} from the scan start to $\Delta t_{AVE}$, the equation:
\begin{equation}
\label{eq:decay_time}
    t_s = t_{a} + \Delta t_{AVE} - \Delta t_{FRT}
\end{equation}
 can be used (confirmed by Siemens Healthineers). The same approach has been also suggested by Paul Kinahan et
al. in the vendor-specific workflow\cite{qiba_suv}.

Philips does not provide a dedicated private attribute for \texttt{Decay Correction Date Time}. Instead, for datasets where \texttt{Decay Correction (0054,1102)} equals \texttt{START}, Philips sets \texttt{Decay Factor (0054,1321)} to a placeholder value of 1.0~\cite{philips_dicom_nuclear_medicine}. Thus, the effective scan start time must be derived using the \texttt{Frame Reference Time (0054,1300)} along with the “average time” \(\Delta t_{AVE}\) for a decaying radionuclide (Eq. \ref{eq:decay_time}).

Bringing together equations \ref{eq:ge_decay_time}, \ref{eq:tave}, \ref{eq:siemens_decay_time}, and \ref{eq:decay_time} allows to derive a general equation for estimation of the scan time:
\begin{equation}
\label{eq:decay_time_all_vendors}
    t_s =
    \begin{cases} 
        t_{a} + \frac{1}{\lambda} \ln \left( \frac{\lambda \Delta t_{AFD}}{1 - e^{-\lambda \Delta t_{AFD}}} \right) - \frac{\ln(\kappa)}{\lambda} & \text{for Siemens Healthineers}, \\[10pt]
        t_{a} - \Delta t_{FRT} & \text{for GE}, \\[10pt]
        t_{a} + \frac{1}{\lambda} \ln \left( \frac{\lambda \Delta t_{AFD}}{1 - e^{-\lambda \Delta t_{AFD}}} \right) - \Delta t_{FRT} & \text{for Philips and} \\ 
    & \text{Siemens Healthineers.}
    \end{cases}
\end{equation}

\subsubsection{PET DICOM Units}
Although BQML is the most common in PET imaging, alternative units also appear in practice. Philips machines can also generate PET data with units of CNTS (counts). To calculate the SUV or the activity concentration (in Bq/mL) in this case, we refer to the equations provided in the Philips DICOM conformance statements \cite{philips_dicom_nuclear_medicine}:

\begin{equation}
\label{eq:Philips_SUV}
\text{SUV} = A'\cdot f_{SUV},
\end{equation}
\begin{equation}
\label{eq:Philips_Act_Concentr}
A = A'\cdot f_{BQML},
\end{equation}
where \(A'\) is defined by Equation \ref{eq:dcm_intensity}, \(f_{SUV}\) is the \texttt{SUV Scale Factor (7053,1000)}, and \(f_{BQML}\) is the \texttt{Activity Concentration Scale Factor (7053,1009)}.

Another commonly used PET DICOM unit is GML (grams per milliliter). Although Siemens Healthineers, GE, and Philips do not mention this unit in their DICOM conformance statements, the DICOM standard documentation \cite{dicom_part03_section_c8_9} indicates that if the unit is GML and the \texttt{SUV Type (0054,1006)} is set to BW (body-weight) or left unspecified, the SUV should be interpreted as a body-weight normalized and equal to $A'$ (Eq. \ref{eq:dcm_intensity}).

\subsection{Validation datasets and Evaluated Software Solutions}
To evaluate the proposed SUV calculation strategy (Fig. \ref{SUV}) and identify potential pitfalls, we tested it using PET datasets avaliable at our institution, as well as data from autoPET challenge \cite{autopet_dataset, gatidis_kuestner_2022}, IBSI II validation data \cite{vallieres_2015_tcia, vallieres_2015_pmb, clark_2013_tcia}, and ACRIN 6685 dataset \cite{Kinahan2019} which include metabolic PET data from Siemens Healthineers, GE, and Philips (Table \ref{dataset}). Our strategy has been implemented in the open-source software Z-Rad \cite{zrad}, which can be used to convert raw PET DICOM files into body-weight normalized SUV NIfTI files. Also, we published our vendor-specific method for SUV calculation on GitHub Gist \cite{GitHubGist}. Additionally, we reviewed third-party software tools, such as PET extention\cite{PETDICOM_Slicer} (version 6840644) for the open-source software 3D Slicer and the commercial software MIM (version 7.4.2). Both solutions are widely used for reading PET DICOM files. This assessment helped us highlight any inconsistencies that may occur in their SUV calculations.

\section{Results}

\subsection{Validation of the Decay Correction Time Estimation}
Vendor-specific private tags were present in 1,301 out of 1,658 BQML PET series from Siemens Healthineers and GE, corresponding to 485,864 out of 578,357 DICOM slices.

For GE data, private tag \texttt{(0009,100d)} was present in 130'717 DICOM slices. Eq.~\ref{eq:ge_decay_time} was validated for all these slices, with no discrepancies observed.

For Siemens Healthineers data, the private tag \texttt{(0071,1022)} was present in 355'147 DICOM slices, allowing validation of both Siemens Healthineers-specific equations (Eq.~\ref{eq:siemens_decay_time} and Eq.~\ref{eq:decay_time}). Eq.~\ref{eq:siemens_decay_time} was accurate within an absolute discrepancy of less than 1.0 s for 95\% of slices, less than 2.0 s for 99.98\% of slices, and did not exceed 3.8 s for the remaining slices. Validation using Eq.~\ref{eq:decay_time} (based on information from Siemens Healthineers) showed an absolute discrepancy below 1.0 s for 99.994\% of slices.

Philips does not provide a private tag for decay correction date and time. Therefore, validation was performed by comparing results from Eq.~\ref{eq:decay_time} against \texttt{Series Time (0008,0031)}. For the studied Philips data, the \texttt{Series Time} closely matched the earlierst \texttt{Acquisition Time} value, suggesting that the series time had not been modified. Validation on 9'136 DICOM slices revealed that the maximum deviation from \texttt{Series Time} was 1.5 s. For 99.1\% of slices, the absolute discrepancy was below 1.0 s. Deviations exceeding 1.0 s occurred only in slices belonging to a single PET series.

\subsection{Comparison of Z-Rad, MIM, and 3D Slicer}

A comparative analysis of SUV values among Z-Rad, MIM, and 3D Slicer for the data in BQML units demonstrated that in 86\% of scans, all three software tools yielded the same SUV values. However, for the remaining 14\% (215 scans), lower SUV values were observed in 3D Slicer and MIM, with an average difference of 32\%. Further investigation revealed that this difference came from the use of the general \texttt{Series Time (0008,0031)} rather than the vendor-specific private tags by 3D Slicer and MIM. In these 215 scans, the \texttt{Series Time (0008,0031)} was shifted by one hour relative to the information stored in the vendor-specific private tag.

MIM and 3D Slicer detected neither when the \texttt{Series Time (0008,0031)} was earlier than the injection time (in 86 cases) nor when it was later than the injection time but still earlier than time in vendor-specific private tag (or than the earliest \texttt{Acquisition Time (0008,0032)}, when no private tag was present), which occurred in 126 cases, as a result producing lower SUV values. This was mostly eliminated by using the vendor-specific private tags \texttt{(0009,100d)} or \texttt{(0071,1022)}. In cases where this private tag was not present, \(t_s\) could be reliably calculated using Eq. \ref{eq:decay_time_all_vendors}. However, in 65.6\% of the BQML data, private tag contained an shifted date compared to the \texttt{Series Date (0008,0021)} and \texttt{Acquisition Date (0008,0022)}, which were consistent. 

For the GML data, MIM handled SUV calculation by following the strategy outlined in the DICOM standard documentation~\cite{dicom_part03_section_c8_9}. In contrast, 3D Slicer applied the QIBA “happy path only” strategy~\cite{qiba_suv_happypath}, resulting in different SUV conversion.

For CNTS data, the 3D Slicer PET module~\cite{slicer_suv_calculator} did not apply the required vendor-specific conversion factors (Eqs.~\ref{eq:Philips_SUV} and \ref{eq:Philips_Act_Concentr}), relying instead on \texttt{Series Time (0008,0031)}, while MIM did not convert PET DICOM files from CNTS to BQML and displayed images only in raw CNTS units.

The studied PET CNTS scans, provided private DICOM attributes enabling accurate SUV calculation via (Eq.~\ref{eq:Philips_SUV}). However, since the \texttt{SUV Type} attribute was not specified, we validated that Eq.~\ref{eq:Philips_SUV} indeed produces SUV values normalized to patient body weight. This was verified for the data where both \texttt{SUV Scale Factor (7053,1000)} and \texttt{Activity Concentration Scale Factor (7053,1009)} were present.

\subsection{Comparison of Proposed and QIBA SUV Calculation Strategies}

When compared to our strategy, the QIBA “happy path only” approach~\cite{qiba_suv_happypath} accounts nither for alterations in time-related DICOM metadata nor for vendor-specific conversion rules. By design, the “happy path only” strategy applies to PET data in BQML units only, providing no support for other unit types. In contrast, our strategy supports vendor-specific conversion rules and BQML, GML, and CNTS units. Due to no additional handling of time-related DICOM metadata by the “happy path only” approach, it led to discrepancies in SUV values of 32\% in 215 PET scans with BQML units from our validation dataset (Table \ref{dataset}).

The vendor-specific QIBA strategy proposed by Paul Kinahan et al.~\cite{qiba_suv_pseudocode} extends the “happy path only” approach by incorporating some vendor-dependent considerations but fails for GE data when early bed positions are missing. This limitation arises because it uses Eq.~\ref{eq:decay_time}, whereas GE data require Eq.~\ref{eq:ge_decay_time}. In addition, this method only addresses cases where \texttt{Series Time (0008,0031)} is later than \texttt{Acquisition Time (0008,0032)}, neglecting instances where the \texttt{Series Time} is earlier. This results in discrepancies identical to those of the QIBA “happy path only” approach — 32\% in 215 PET scans—whereas our strategy is directly based on vendor-specific metadata and remains robust to general DICOM attributes alternations.

\section{Discussion}

Both QIBA strategies proposed by Paul Kinahan et al.~\cite{qiba_suv_happypath, qiba_suv_pseudocode} do not fully account for vendor-specific details and alterations in time-related DICOM metadata, leading to discrepancies in SUV values of up to 32\% in the studied dataset. Therefore, it is recommended to use the time component indicated in private tags whenever possible, and when these tags are absent, to calculate the scan start time based on \texttt{Acquisition Time (0008,0032)}, \texttt{Decay Factor (0054,1321)}, \texttt{Actual Frame Duration (0018,1242)}, and \texttt{Frame Reference Time (0054,1300)} as shown in Eq. \ref{eq:decay_time_all_vendors}. The proposed strategy combines vendor-specific attributes with fallback calculations based on general DICOM attributes when necessary, this method provides a robust and transparent solution to a persistent challenge in quantitative PET imaging. During the vendor-specific validation of the decay correction time estimation, we did not observe differences between models originating from the same vendor, indicating that all included vendors were consistent in their strategies across different PET scanner models.

Unfortunately, we were not able to obtain PET data from  other vendors  (e.g.,  Canon, United Imaging), thus limiting validation of the current strategy to scanners from Philips, GE, and Siemens Healthineers. Future work will include maintaining the proposed strategy and updating it according to vendor-specific DICOM conformance statements, as well as incorporating additional PET hardware vendors. 

The proposed strategy is implemented in publicily available and open-source software Z-Rad \cite{zrad}. Additionally, we provide lightweight Python code for SUV calculation on GitHub Gist \cite{GitHubGist}.

\section{Conclusions}

Our proposed vendor-specific SUV calculation strategy substantially improves PET quantitative analyses by accommodating vendor-specific DICOM attributes and correcting for discrepancies in \texttt{Series Time (0008,0031)} and \texttt{Acquisition Time (0008,0032)}. These adjustments resolve the common pitfalls associated with the QIBA approaches, particularly in cases where \texttt{Decay Correction Date Time} attributes are either missing or misaligned.

\section{Acknowledgments}
This study was supported by Comprehensive Cancer Center Zurich (C3Z Precision Oncology Funding Program, OMD project). We would also like to thank GE and Siemens Healthineers for their helpful input regarding SUV definitions specific to their respective systems.

\section{Conflict of Interest}
The department of Radiation Oncology, University Hospital Zurich has teaching and research agreements with Siemens Healthineers.
Stephanie Tanadini-Lang received travel funds from Siemens Healthineers. The remaining authors declare no conflict of
interest.

\section{Figures and Tables}
\begin{figure}[ht]
   \begin{center}
   \includegraphics[width=14cm]{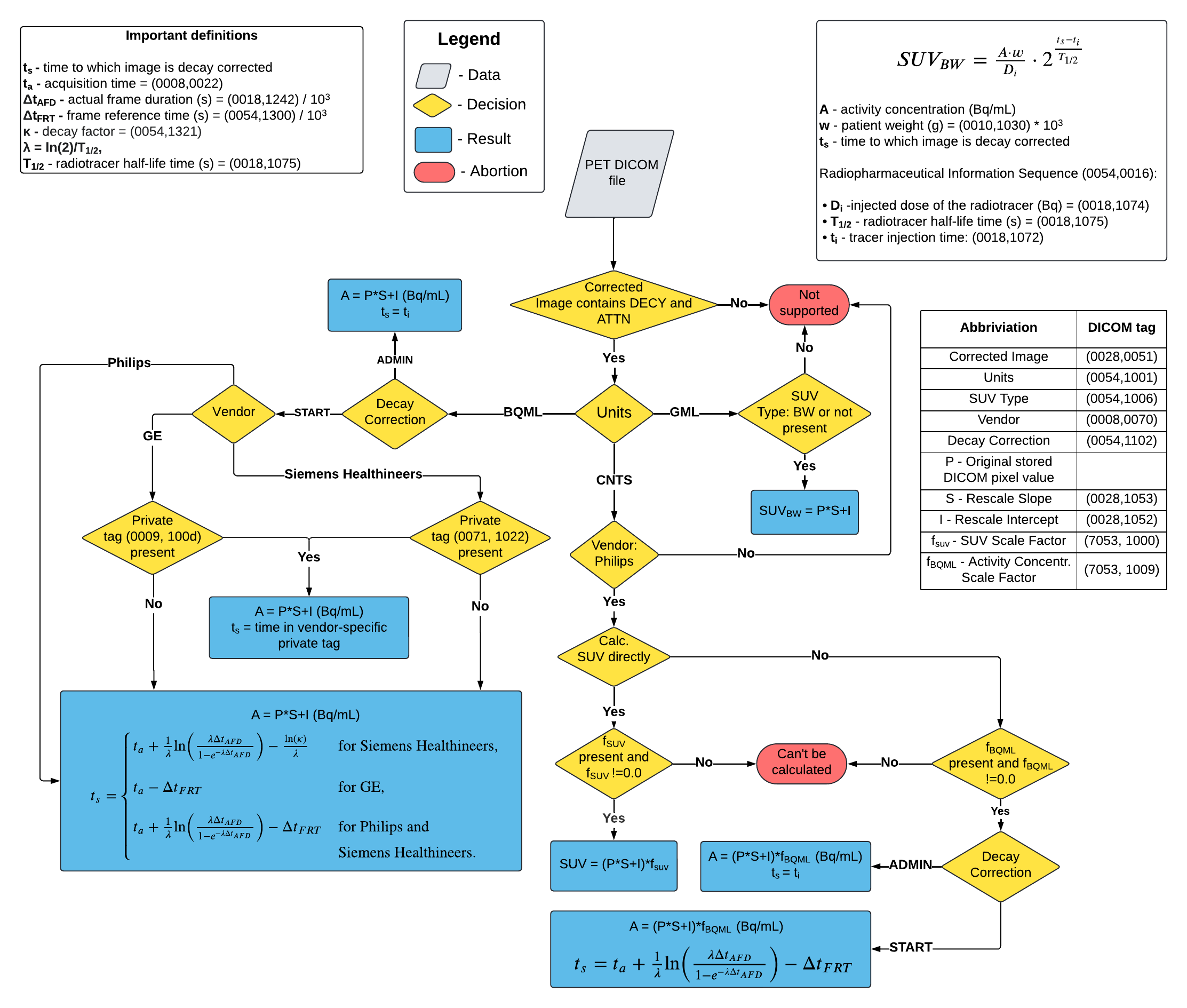}
   %
   %  conclusion, use directly created .pdf from agr for best results.
   %
   \caption{Vendor-specific SUV calculation strategy.}
   \label{SUV}  %note label inside caption
    \end{center}
\end{figure}

\begin{table}[htbp]
\begin{center}
\caption{Vendor-specific SUV calculation strategy.
\label{dataset}
\vspace*{2ex}
}
\begin{tabular} {|l|p{2.5cm}|p{4cm}|p{3cm}|c|}
\hline
Data Source & Vendor  & Manufacturer Model Name (0008,1090) & Units (0054,1001)& No. of PET Series \\
\hline
autoPET II \cite{autopet_dataset, gatidis_kuestner_2022}   & Siemens Healthineers & Biograph mCT, mCT Flow, Biograph 64& BQML (100\%)& 1014 \\
\hline
\begin{tabular}[t]{@{}l@{}}External \\ and internal data\end{tabular} & GE & Discovery MI, Discovery 690, Discovery STE, Discovery 710, Discovery RX &BQML (100\%)& 522 \\
\hline
\begin{tabular}[t]{@{}l@{}}External data, \\ ACRIN 6685\cite{Kinahan2019} \end{tabular} & Philips &Allegro Body, Guardian Body, GEMINI TF 16, GEMINI TF 64, GEMINI TF Big Bore, or not specified& BQML(70\%), CNTS (30\%)& 71 \\
\hline

IBSI II \cite{vallieres_2015_tcia, vallieres_2015_pmb, clark_2013_tcia} & GE & Discovery ST & GML (100\%)& 51  \\
\hline

\end{tabular}
\end{center}
\end{table}

% following only if there is an appendix

\newpage

\section*{Appendix}
\addcontentsline{toc}{section}{\numberline{}Appendix}

\begin{longtable}{|p{3.5cm}|p{2cm}|p{8cm}|}
\caption*{DICOM attributes used for SUV calculation.} \\
\hline
\textbf{Attribute Name and Tag} & \textbf{Vendor-specific} & \textbf{Attribute Description} \\
\hline
\endfirsthead

\hline
\textbf{Attribute Name and Tag} & \textbf{Vendor-specific} & \textbf{Attribute Description} \\
\hline
\endhead

\hline \multicolumn{3}{|r|}{\textit{Continued on next page}} \\
\hline
\endfoot

\hline
\endlastfoot

Rescale Intercept (0028,1052) & No & The value b in relationship between stored values (P) and the output units. Units = m*P + b. \\
\hline
Rescale Slope (0028,1053) & No & m in the equation specified by Rescale Intercept (0028,1052). \\
\hline
Units (0054,1001) & No & The units of the pixel values obtained after conversion from the stored pixel values (P) to pixel value units, as defined by Rescale Intercept (0028,1052) and Rescale Slope (0028,1053). \\
\hline
Corrected Image (0028,0051) & No & One or more values that indicate which, if any, corrections have been applied to the images in this Series. \\
\hline
Manufacturer (0008,0070) & No & Manufacturer of the equipment that produced the Composite Instances. \\
\hline
Decay Correction (0054,1102) & No & The real-world event to which images in this Series were decay corrected. \\
\hline
Acquisition Time (0008,0032) & No & The time the acquisition of data that resulted in this image started. \\
\hline
Series Time (0008,0031) & No & Time the Series started. \\
\hline
SUV Type (0054,1006) & No & Type of Standardized Uptake Value (SUV). \\
\hline
Patient's Weight (0010,1030) & No & Weight of the Patient, in kilograms. \\
\hline
Radionuclide Total Dose (0018,1074) & No & The radiopharmaceutical dose administered to the patient measured in Becquerels (Bq) at the Radiopharmaceutical Start Time (0018,1072). \\
\hline
Radiotracer half-life time (0018,1075) & No & The radionuclide half life, in seconds, that was used in the correction of this image. \\
\hline
Tracer injection time: (0018,1072) & No & Time of start of administration. The actual time of radiopharmaceutical administration to the patient for imaging purposes, using the same time base as Series Time (0008,0031). \\
\hline
Decay Factor (0054,1321) & No & The decay factor that was used to scale this frame. \\
\hline
Frame Reference Time (0054,1300) & No &  The time that the pixel values in the image occurred. Frame Reference Time is the offset, in msec, from the Series reference time. \\
\hline
Actual Frame Duration (0018,1242) & No &  Elapsed time for data acquisition in msec. \\
\hline
SUV Scale Factor (7053,1000) & Philips & This value only applies when Units (0054,1001) is equal to CNTS. The SUV Scale Factor is used to convert the pixel data from counts to an SUV value. This is done by using the formula SUV Value = ((P * m) + b) * f, where: P - original stored pixel value, m - Rescale Slope (0028,1053), b - Rescale Intercept (0028,1052), f - SUV Scale Factor (7053,1000). 

If the SUV Scale Factor is 0.0, then the pixel data cannot be converted from counts to an SUV value. \\
\hline
Activity Concentration Scale Factor (7053,1009) & Philips &  This value only applies when Units (0054,1001) is equal to CNTS. The Activity Concentration Scale Factor is used to convert the
pixel data from counts to Activity Concentration (in Bq/mL). This is done by using the formula Activity Concentration Value = ((P *
m) + b) * f, where: P - original stored pixel value, m - Rescale Slope (0028,1053), b - Rescale Intercept (0028,1052), f - Activity Concentration Scale Factor (7053,1009)

If the Activity Concentration Scale Factor is 0.0, then the pixel data cannot be converted from counts to Activity Concentration.
\\
\hline
Decay Correction DateTime (0071,1022) & Siemens Healthineers &  - \\
\hline
Decay Correction DateTime (0009,100d) & GE & -\\
\hline
\end{longtable}

%\section*{References}
%\addcontentsline{toc}{section}{\numberline{}References}
%\vspace*{-20mm}

% Following assumes you are using bibtex. However, for submission to the
% journal you MUST explicitly INCLUDE THE REFERENCES IN THE TEX FILE. 
% In that case you need the following

%\begin{thebibliography}{10}
% insert the .bbl file generated by bibtex here
	%This will be a series of entries from your .bib file formatted
	%something like
	%\bibitem{Me09}
        %{I.~Meijsing, B.~W.~Raaymakers, A.~J.~E.~Raaijmakers \it et al.},
        %\newblock {Dosimetry for the MRI accelerator: the impact of a 
	%magnetic field on the response of a Farmer NE2571 ionization chamber},
        %\newblock Phys. Med. Biol. {\bf 54}, 2993 -- 3002 (2009).

% \end{thebibliography}

% The following is when using bibtex and picks up the example.bib file

%\bibliography{Explicit address of .bib file}
\bibliography{references.bib}      %example.bib is on the same directory
% above points to where we find the master reference list
% and also causes the bibliography to be printed

% When creating your bibliography you should run bibtex on your local
% computer after running pdflatex on your .tex file. bibtex will
% generate a .bbl file.
% Copy the contents of this .bbl file into your main latex document,
% replacing the "\bibliography" command which was pointing at your .bib file.

% following defines style of .bbl file 

%\bibliographystyle{explicit relative path to medphy.bst}
\bibliographystyle{medphy.bst}    %if this is installed on your system,
				    %it is not essential to have the    ./

% Note that you need to typeset once, then run bibtex, then typeset another
% two times to get the references working properly.

\end{document}